\documentclass[
 superscriptaddress,
 amsmath,amssymb,
 reprint,
]{revtex4-2}

\usepackage{pgfplots}
\usepackage{booktabs}
\usepackage{dcolumn}
\usepackage[colorlinks=true, citecolor=blue, urlcolor=blue,  linkcolor=blue]{hyperref}

\newcommand{\abs}[1]{\left\lvert#1\right\rvert}
\newcommand{\avg}[1]{\left\langle#1\right\rangle}
\newcommand{\centered}[1]{\begin{tabular}{l} #1 \end{tabular}} 
\let\originalleft\left
\let\originalright\right
\renewcommand{\left}{\mathopen{}\mathclose\bgroup\originalleft}
\renewcommand{\right}{\aftergroup\egroup\originalright}

\begin{document}


\title{On the topology of the space of coordination geometries}

\author{John \c{C}amk{\i}ran}
\email{john.camkiran@utoronto.ca}
\affiliation{\mbox{Department of Materials Science and Engineering, University of Toronto, Toronto, Ontario, M5S 3E4, Canada}}

\author{Fabian Parsch}
\affiliation{\mbox{Department of Mathematics, University of Toronto,  Toronto, Ontario, M5S 2E4, Canada}
}
\affiliation{\mbox{Department of Materials Science and Engineering, University of Toronto, Toronto, Ontario,  M5S 3E4, Canada}}

\author{Glenn D. Hibbard}
\affiliation{\mbox{Department of Materials Science and Engineering, University of Toronto, Toronto, Ontario, M5S 3E4, Canada}}

\date{\today}

\begin{abstract}
Coordination geometries describe how the neighbours of a central particle are arranged around it. Such geometries can be thought to lie in an abstract topological space; a model of this space could provide a mathematical basis for understanding physical transformations in crystals, liquids, and glasses. With this motivation, the present work proposes a metric model of the space of three-dimensional coordination geometries. This model is conceived through the generalisation of a local orientational order parameter and seems to be consistent with geometric intuition. It appears to suggest a taxonomy of coordination geometries with five main classes, each with a distinct character. A quantitative notion of orientational typicality is introduced and its interplay with orientational order is found to evidence a statistical regularity with respect to point symmetry. By the assertion of axioms on the topology of the space herein modelled, the range of structures that are possible to resolve with the order parameter in molecular dynamics simulations is greatly increased.
\end{abstract}

\maketitle


\section{Introduction}

In both crystalline and noncrystalline materials, the geometry of the immediate surrounding of a particle often undergoes transformation between seemingly dissimilar geometries of coordination \cite{moroni_2005, natarajan_2018, plessow_2020, hu_2022}. A scientific understanding of precisely which transformations are possible is precluded by the lack of a mathematical model of the abstract topological space that coordination geometries can be thought to lie in. Prior discussion on such a space is limited \cite{terrones_1994, kolli_2020, thomas_2021}, presumably because there is little interesting to be said about it from the traditional perspective of symmetry.

\begin{figure}[b]
    \centering
    \includegraphics[width=\linewidth]{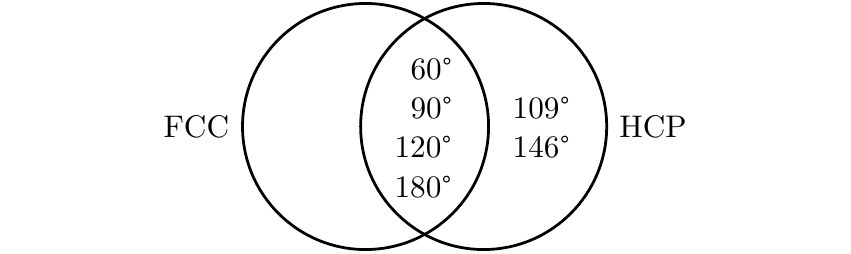}
    \caption{Bond angles for the FCC and HCP coordination geometries, rounded to the nearest integer. The presence of many common bond angles suggests a strong similarity between the geometries. Such is not obvious from their one-particle coefficients, $E_\text{FCC} \approx 4.04$ and $E_\text{HCP} \approx 3.46$.}
    \label{fig:venn}
\end{figure}

In a recent publication, the authors of this work discussed a local structural phenomenon called \textit{extracopularity}---the tendency of particles in condensed phases to have far fewer distinct bond angles than combinatorially possible \cite{camkiran_2022}. Extracopularity is an informational redundancy that accompanies local orientational order and is quantified by the \textit{extracopularity coefficient} $E$, an order parameter in the sense of the work by Steinhardt and colleagues \cite{steinhardt_1983}. Being a strictly local quantity, the original extracopularity coefficient says little about the relationships between coordination geometries. Fig. \ref{fig:venn} illustrates this point.

To enable the study of these relationships, the present work generalises $E$ to $n$ particles. The generalised quantity is a statistical expansion of the original one and thus enjoys a similar, information-theoretic interpretation. A metric for three-dimensional coordination geometries follows naturally from the generalisation. The space endowed with this metric exhibits clustering around geometries that are similar in construction, thereby validating it as a model of the hypothetical abstract space of coordination geometries. But many of the remaining features of this space are nontrivial, making it a possible new source of theoretical insight on the atomic-scale structure of crystals, liquids, and glasses \cite{pauling_1929, kolli_2020, thomas_2021}.

The main computational challenge with a metric based on  $n$-particle $E$ is that evaluating the latter quantity requires the explicit discretization of bond angles. Such was avoidable with one-particle $E$ \cite{camkiran_2022}. 
Here we present a method of bond angle discretization that is in a certain sense optimal. This method is based on the observation that bond angles are more likely to take certain values than others. The method can also be used to compute one-particle extracopularity coefficients with greater precision than previously possible. It therefore has immediate practical implications on structural analysis in the molecular dynamics setting.

The remainder of this work is organised as follows: \mbox{Sec. \ref{sec:generalization}} reviews $E$ and discusses its generalisation to $n$ particles. \mbox{Sec. \ref{sec:metric}} proposes a metric for three-dimensional coordination geometries based on this generalisation. \mbox{Sec. \ref{sec:discretization}} addresses the computational problem of bond angle discretization. \mbox{Sec. \ref{sec:topology}} studies the topology of the space of three-dimensional coordination geometries. Finally, \mbox{Sec. \ref{sec:discussion}} discusses a few implications of our results.

\section{Generalisation}
\label{sec:generalization}

We begin this first section by briefly reviewing the one-particle extracopularity coefficient. We then make precise the behaviour that is desired of its $n$-particle generalisation. Lastly, we define such a generalisation, obtain its closed form, and check that it behaves in the desired way.

\subsection{Review}

Consider a finite nonempty set $S$ of points in Euclidean space, and call each point $p \in S$ a \textit{particle}. Define the \textit{neighbourhood} $\mathcal{N}(p)$ of a particle $p$ as a subset of $S$ not containing $p$, and call its elements $q \in \mathcal{N}(p)$ the \textit{neighbours} of $p$. Given a particle $p$ and its any neighbour $q$, call the vector $q-p$ a \textit{bond}. Typically, the neighbourhood of a particle is not known \textit{a priori} and must be determined empirically. In particle packings, this can be achieved through the robustified Voronoi tessellation \cite{camkiran_2022}. 

Let $B_i$ denote the set of all unordered bond pairs for the  $i$th particle and $\Theta_i$ its set of (smaller) pairwise bond angles. Finally, let $I(A) = \log_2\abs{A}$ give the \textit{information content} of a discrete set $A$ \cite{hartley_1928}. Then, the (one-particle) extracopularity coefficient $E_i$ of the $i$th particle is defined by the following difference:
\begin{equation}
    E_i= I(B_i) - I(\Theta_i).
\end{equation}

Information-theoretically, this quantity tells us how much easier it would be to search for a specific bond pair if the angle made by the pair were known.

For a particle with $k_i>1$ bonds and $m_i= \abs{\Theta_i}$ distinct bond angles, $E_i$ can be written more explicitly as follows:
\begin{equation}
    E_i= \log_2 \left[ \frac{k_i^2-k_i}{2m_i} \right].
    \label{eq:original}
\end{equation}
We now consider a generalisation of this quantity to $n$ particles.

\subsection{Desideratum}
\label{sec:desideratum}

The simple fact that some quantity generalises the extracopularity coefficient to $n$ particles does not guarantee that it will be useful. Hence, before defining such a quantity, we must carefully consider the behaviour needed to make it so.

Let $E_{1 \dots n}$ denote the extracopularity coefficient of a nonempty collection of particles. If each of these particles were to have the same coordination geometry, it would only be natural for the coefficient of the collection to be equal to the coefficient corresponding to that geometry. More concretely, if these $n$ particles are arranged, say, as an FCC crystal, then one would expect to have $E_{1 \dots n} = E_\text{FCC} \approx 4.04$. Furthermore, it would not make sense for a particle collection to be more ordered as a whole than it is locally around any of its constituent particles. We capture the above behaviour in the following upper bound:
\begin{equation}
    E_{1 \dots n} \leq \max \{E_1, \dots, E_n\},
    \label{eq:desideratum}
\end{equation}
with equality if and only if $(k_i, \Theta_i) = (k_j, \Theta_j)$ for all $i, j$.

Observe that a necessary condition for this bound to hold with equality is that $E_i= E_{j}$ for all $i, j$. Notice, however, that this condition is not sufficient, since the fraction in the argument of the logarithm in \mbox{Eq. (\ref{eq:original})} is not unique to every pair $(k_i,\Theta_i)$.

\subsection{Definition}

Having formalised the desired behaviour, we are ready to devise a generalisation able to satisfy it. Define the \textit{extracopularity coefficient} ${E_{1 \dots n}}$ of a collection of $n$ particles by
\begin{equation}
    E_{1 \dots n} = \avg{I\left(B_{i}\right)} - I(\Theta_{1 \dots n}),
    \label{eq:definition}
\end{equation}
where $\avg{\cdot}$ denotes the arithmetic average and $\Theta$ denotes the set of all bond angles exhibited by the particles.

Much like the one-particle coefficient $E_i$, the $n$-particle coefficient $E_{1 \dots n}$ can be expressed more explicitly in terms of the number of bonds and distinct bond angles. From elementary combinatorics, we have
\begin{align}
    I(B_{i}) &= \log_2 \abs{B_{i}}  \\
    &= \log_2 \left[\frac{k_{i}^2-k_{i}}{2} \right].
\end{align}
Taking the arithmetic average of this quantity gives us the first term in the definition of $E_{1 \dots n}$,
\begin{align}
    \avg{I\left(B\right)} &= \frac{1}{n} \sum_{i=1}^n\log_2 \left[\frac{k_i^2-k_i}{2} \right] \\
    &= \frac{1}{n} \log_2 \prod_{i=1}^n\left[\frac{k_i^2-k_i}{2} \right]  \\
    &= \frac{1}{n} \log_2 \left[ \frac{\prod\limits_{i=1}^n\left(k_i^2-k_i\right)}{2^n} \right] \\
    &= \log_2 \left[ \frac{ \sqrt[\leftroot{-2}\uproot{2}n]{\prod\limits_{i=1}^n\left(k_i^2-k_i\right)} }{2} \right].
    \label{eq:term-1}
\end{align}
The second term in the definition can be expanded as follows:
\begin{align}
    I\left(\Theta_{1 \dots n}\right) &= \log_2 \abs{\Theta_{1 \dots n}} \\
    &= \log_2 \abs{ \bigcup_{i=1}^n\Theta_i}.
    \label{eq:term-2}
\end{align}
Combining Eq. (\ref{eq:definition}), (\ref{eq:term-1}), and (\ref{eq:term-2}) reveals a general formula for the extracopularity coefficient of $n$ particles,
\begin{equation}
    E_{1 \dots n} = \log_2\left[ \frac{ \sqrt[\leftroot{-2}\uproot{2}n]{\prod\limits_{i=1}^n\left(k_i^2-k_i\right)} }{2 \abs{ \bigcup\limits_{i=1}^n\Theta_i} } \right].
    \label{eq:formula}
\end{equation}
Observe that taking $n=1$ recovers Eq. (\ref{eq:original}), showing that what we have obtained is indeed a generalisation of the one-particle case. The focus of the present work is on the case of $n=2$.

\subsection{Bounds}

While clearly desirable, the upper bound discussed in \mbox{Sec. \ref{sec:desideratum}} is difficult to prove, as doing so would require a precise understanding of the relationship between bond angles and the number of bonds. There are nonetheless certain indications that it may be satisfied. Notably, the following looser bound is found to hold:
\begin{equation}
    E_{1 \dots n} \leq \log_2 \left[ \frac{\max_i(k_i^2 -k_i)}{2\min_i(m_i)} \right],
    \label{eq:loose-bound}
\end{equation}
with equality if and only if $(k_i, \Theta_i) = (k_j, \Theta_j)$ for all $i, j$.

Let us show that this is indeed the case. The argument of the logarithm in the formula for ${E_{1 \dots n}}$ is a positive fraction, which increases with its numerator and decreases with its denominator. Observe that the numerator of the argument is simply the geometric average of $k_i^2-k_i$. It is known that a geometric average cannot be larger than the largest of the numbers being averaged,
\begin{equation}
    \sqrt[\leftroot{-2}\uproot{2}n]{ x_1 \dots x_n } \leq \max  \{ x_1, \dots, x_n \},
\end{equation}
with equality if and only if $x_i = x_{j}$ for all $i,j$. Next observe that the  denominator of the argument is twice the cardinality of a union. Certainly, the cardinality of a union cannot be less than the cardinality of the sets under union,
\begin{equation}
    \abs{\bigcup_{i=1}^n A_i} \geq \min\{ \abs{A_1},\dots, \abs{A_n} \},
\end{equation}
with equality if and only if $A_i$ are identical. Eq. (\ref{eq:loose-bound}) is now immediate.

For completeness, let us also consider the corresponding lower bound. A geometric average cannot be smaller than the smallest of the numbers being averaged,
\begin{equation}
    \sqrt[\leftroot{-2}\uproot{2}n]{ x_1 \dots x_n } \geq \min  \{ x_1, \dots, x_n \}.
\end{equation}
And the cardinality of a union is no larger than the sum of the cardinalities of the sets under union,
\begin{equation}
    \abs{\bigcup_{i=1}^n A_i} \leq \sum_{i=1}^n \abs{A_i}.
\end{equation}
Thus, ${E_{1 \dots n}}$ is bounded from below as follows:
\begin{equation}
    E_{1 \dots n} \geq \log_2 \left[ \frac{\min_i(k_i^2 -k_i)}{2(m_1 + \dots + m_n)} \right].
\end{equation}

\section{Metric}
\label{sec:metric}

The relationship between coordination geometries is of both practical \cite{stukowski_2012, tanaka_2019} and theoretical \cite{yang_2014, thomas_2021} interest. Prior efforts to capture these relationships tend to have done so through general dissimilarity functions \cite{terrones_1994, hundt_2006, yang_2014}. Such functions, however, lack one or more of the properties that are important to quantifying the degree of dissimilarity between two objects \cite{osearcoid_2006}. Much more informative are \textit{metric} dissimilarity functions, often called distance functions or metrics. Given a set $M$, a map $d:M \times M \to \mathbb{R}$ is said to be a \textit{metric} on $M$ if it satisfies the following three properties:
\begin{equation}
    \begin{aligned}
    \text{I.} \quad \null & d(x,y) = 0 \quad \text{if and only if} \quad x = y \\
    \text{II.} \quad \null &d(x,y) = d(y,x) \\
    \text{III.} \quad \null &d(x,z) \leq d(x,y) + d(y,z)
    \end{aligned}
    \label{eq:metric-properties}
\end{equation}

Here we seek such a function on the set $G$ of three-dimensional coordination geometries. Given two geometries $g,h \in G$, define the \textit{extracopularity distance} $d_E(g,h)$ between them by
\begin{equation}
    d_E(g,h) = \max \{E_{g},E_{h}\} - E_{gh},
    \label{eq:extracopularity-distance}
\end{equation}
where $E_{g}$ is understood to be the extracopularity coefficient of a particle with coordination geometry $g$ and $E_{gh}$ that of a pair of particles with geometries $g$ and $h$. \mbox{Property II} is immediate from the symmetries of the set maximum and the two-particle extracopularity coefficient. While Properties I and III are difficult to prove analytically, they can  be shown numerically for $22$ of the most important geometries of coordination, which are given in \mbox{Appendix \ref{appx:cegs}}.

\section{Bond angle discretization}
\label{sec:discretization}

As visible from Eq. (\ref{eq:extracopularity-distance}), the extracopularity distance is ultimately the difference between a one-particle and a two-particle extracopularity coefficient. The key to computing both kinds of coefficients is in determining the number of \textit{distinct} bond angles $\abs{\Theta}$ for the particle(s) being studied. This is typically much lower than the naive count of bond angles due to equivalences between angles. For example, the three instances of the $180^\circ$ degree bond angle class in a simple cubic neighbourhood may be measured as $176.4^\circ$, $178.7^\circ$, and $179.2^\circ$. Thus, a naive count of these angles would result in overcounting $180^\circ$ by two. Clearly, each instance of an angle must be mapped to the angle class before an informative count can take place. We call this process \textit{bond angle discretization}. In this section, we discuss a bond angle discretization method that is optimal in a specific sense. Previously, one-particle $E$ had been computed without discretization using a workaround that resulted in lower precision. An updated version of our algorithm that uses the discretization method discussed below is publicly available \footnote{Code available at \url{www.github.com/johncamkiran/extracopularity}}.

\subsection{Approach}

The most straightforward approach to bond angle discretization is what one might call \textit{fixed discretization}, wherein the interval $(0,180]$ is partitioned into subintervals called \textit{bins}, each having fixed endpoints, called \textit{edges}. Such an approach is justified by the observation that bond angles do not take values in $(0,180]$ with uniform probability. Observe that the repulsive forces that prevent (physical) particles from getting too close to each other also make certain angles more likely than others. The angle of $60^\circ$ for instance is quite common, as it is observed whenever three particles all neighbour each other with equal radial distance. In this way, $60^\circ$ can be seen as an \textit{inherent} bond angle of systems of interacting particles. Given a comprehensive list of such bond angles, fixed discretization can be performed optimally by placing bin edges at the points in $(0,180]$ that lie between them. A technicality of this approach is discussed in Appendix \ref{appx:discretization}.

\subsection{Method}

We now outline a method of bond angle discretization with the fixed approach. We begin by listing the bond angles exhibited by particles with commonly encountered geometries of coordination (given in \mbox{Appendix \ref{appx:cegs}}). We omit a geometry if any other geometry can be described as a capping (or augmentation) of it. For instance, we include BSA but not SA and CSA, as the former constitutes a capping of the latter. This is done to avoid biasing the bond angle list towards clusters of geometries that are essentially identical in construction. Then, inherent angles are taken to be the points in $(0, 180]$ that are locally maximal in bond angle density ($0$ is taken to be inherent by convention). We establish these points using the algorithm DBSCAN \cite{ester_1996}, which takes the following two arguments: the minimum number of nearby points $\mathrm{minPts}$ needed for a point to be considered an inherent angle candidate, and the search radius $\varepsilon$ that defines `nearby'.

We select the first argument $\mathrm{minPts} = 1$, which is the most conservative choice as it does not disqualify any point from inherent angle candidacy. The choice of the second argument $\varepsilon$ is more subtle. Visibly, larger values of $\varepsilon$ will lead to coarser and hence more statistically robust discretisations. If chosen too large, however, substantially different angles will be equivalated, rendering the result of the discretization uninformative. To optimise this choice, we assert two axioms regarding the topology of the space of coordination geometries:

\begin{enumerate}
    \item \begin{enumerate}
        \item \textit{FCC is closer to HCP than it is to BCC.}
        \item \textit{HCP is closer to FCC than it is to BCC.}
    \end{enumerate}
    \item \begin{enumerate}
        \item \textit{CSA and BSA are the two closest to SA.}
        \item \textit{CSP and BSP are the two closest to HDR.}
    \end{enumerate}
\end{enumerate}

The first axiom simply formalises the well-known intimate link between the constructions of FCC and HCP. The second axiom is justified by the fact that capping constitutes only a minimal change to any geometry. We thus arrive at our choice of $\varepsilon = 2.85$, the largest value to two decimal places for which the above axioms are satisfied.

\section{The topology}
\label{sec:topology}

\begin{figure}[b]
    \centering
    \includegraphics[width=\linewidth]{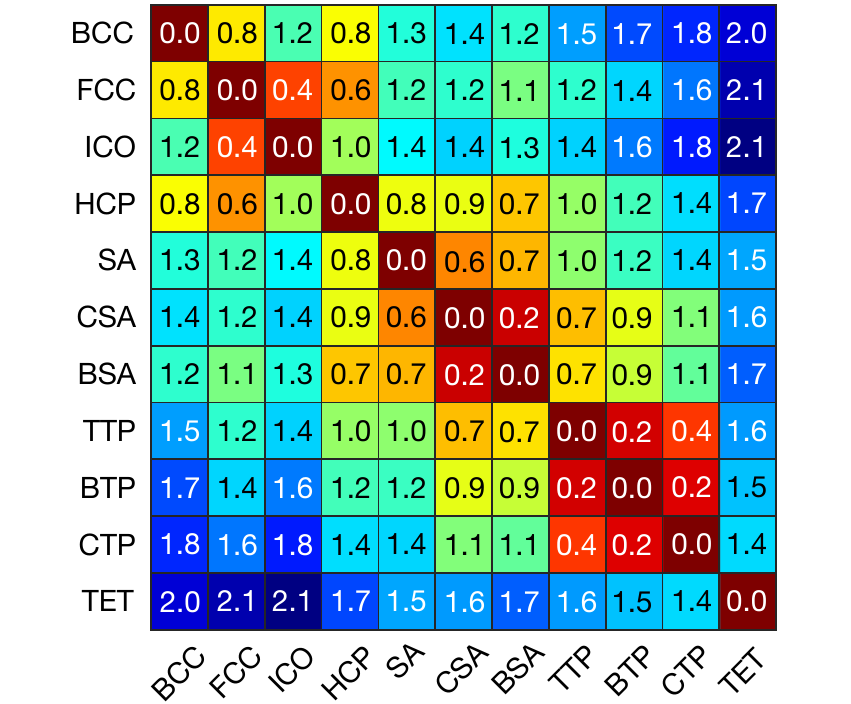}
    \caption{Heatmap of extracopularity distances for $11$ of the $22$ commonly encountered geometries (see Appendix \ref{appx:cegs} for the full version). All values are rounded to one decimal place. Rows/columns are ordered as per \mbox{Ref. \cite{bar-joseph_2001}}.}
    \label{fig:heatmap-part}
\end{figure}

To evaluate our model numerically, we considered $22$ of the most commonly encountered geometries of coordination. We computed all $231$ extracopularity distances between these geometries, given partially in Fig. \ref{fig:heatmap-part} and fully in Fig \ref{fig:heatmap-full}. These distances were found to satisfy the properties of a metric [Eq. (\ref{eq:metric-properties})], and the extracopularity coefficients they were computed from were found to satisfy the desired upper bound [Eq. (\ref{eq:desideratum})]. We employed two common techniques to make sense of this distance data, namely hierarchical clustering and multidimensional scaling.

\subsection{Hierarchical clustering}

\begin{figure}[b]
    \centering
    \includegraphics[width=\linewidth]{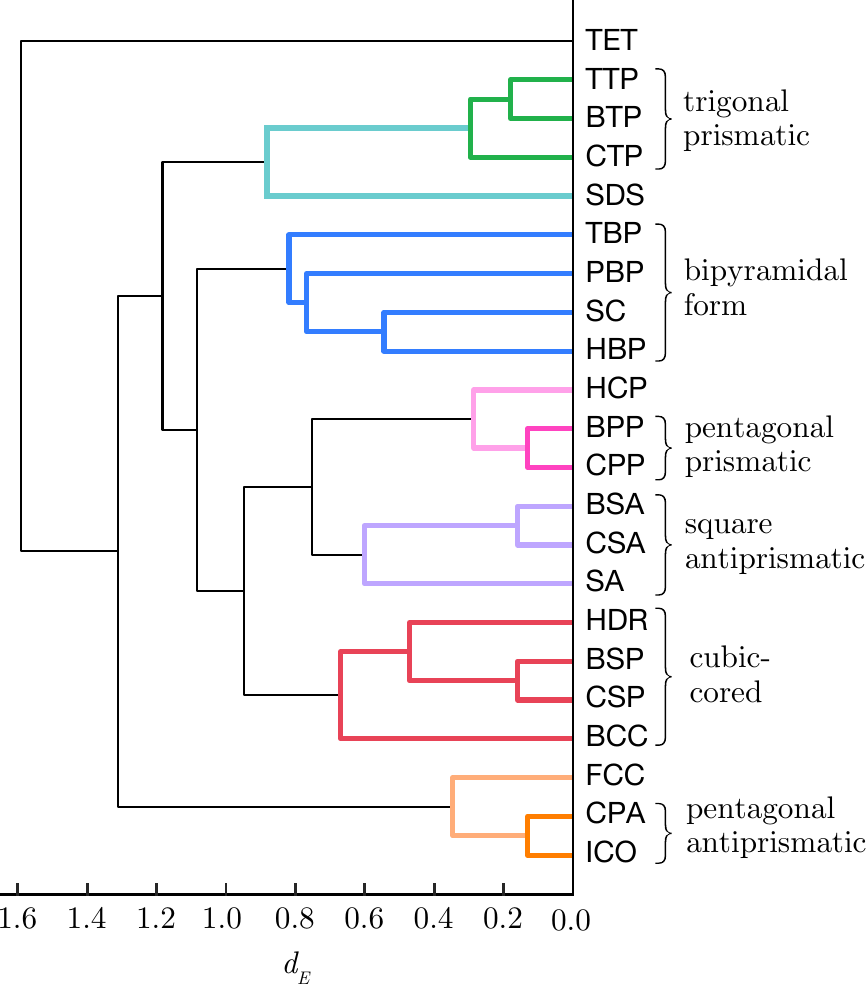}    \caption{Dendrogram of the binary tree obtained through the hierarchical clustering of commonly encountered geometries based on $d_E$. The height of each branch (as measured horizontally) corresponds to the average extracopularity distance between the geometries that it connects. Leaves are ordered as per \mbox{Ref. \cite{bar-joseph_2001}}.}
    \label{fig:dendrogram}
\end{figure}

Hierarchical clustering is the construction of a binary tree through the iterative pairing of a set of items given the distances between them \cite{sneath_1973}. Each iteration pairs the two items that are closest to each other, which are thereafter treated as a single item. The distance to/from this new item is calculated as the average distance to/from its constituents. The iteration stops when only one item remains. Fig. \ref{fig:dendrogram} illustrates the result of hierarchical clustering for commonly encountered coordination geometries based on the extracopularity distance.

The tree was found to exhibit clustering around collections of geometries with similar construction, namely those pentagonal prismatic, square antiprismatic, trigonal prismatic, bipyramidal, cubic-cored, and pentagonal antiprismatic. Among the less expected pairings were that of HCP with the pentagonal prismatics, FCC with the pentagonal antiprismatics, and SDS with the trigonal prismatics. The technique we discuss next offers some perspective on these pairings.

\subsection{Multidimensional scaling}

\begin{figure}[t]
    \centering
    \includegraphics[width=\linewidth]{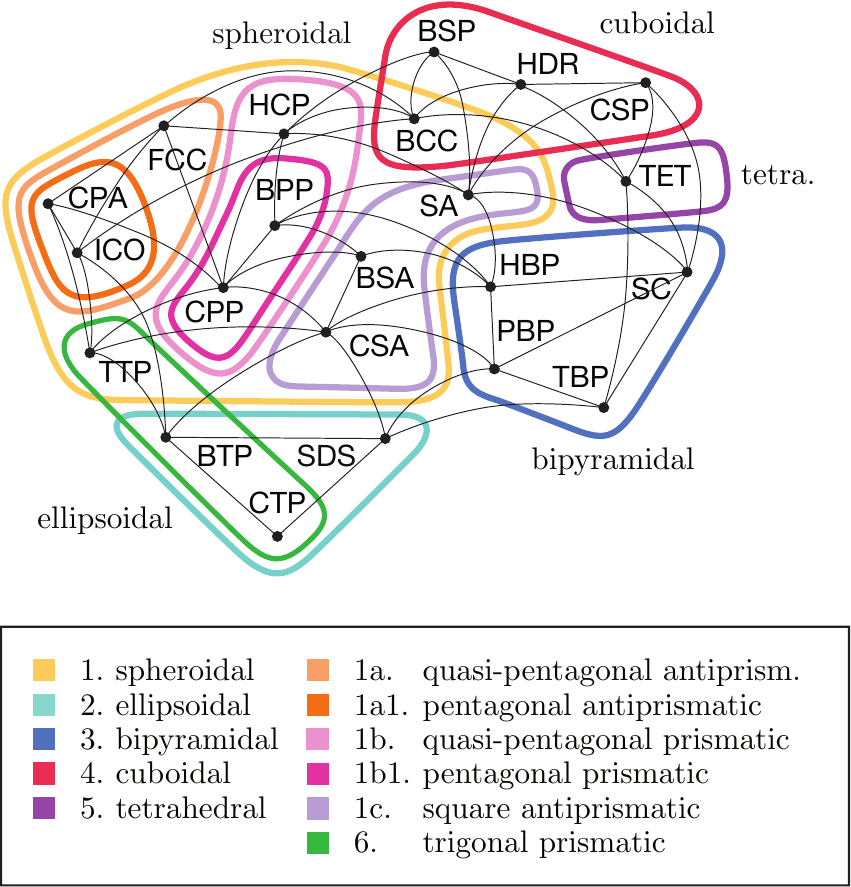}
    \caption{Graph of commonly encountered geometries implied by the Delaunay triangulation of the two-dimensional scaling of extracopularity distances. Edge lengths have no significance.}
    \label{fig:graph}
\end{figure}

Multidimensional scaling (MDS) is the practice of embedding a set of items into an abstract Cartesian space given their pairwise distances \cite{cox_1994}. We used nonclassical MDS with metric stress, which is the algorithm appropriate for non-Euclidean metrics. We established $8$ as a sufficient number of dimensions for the scaling, as additional dimensions were not found to lead to further reductions in stress. Fig. \ref{fig:graph} depicts the graph implied by the Delaunay triangulation \cite{devadoss_2011} of the two-dimensional approximation of the full, 8-dimensional scaling.

\subsubsection{Taxonomy}
\label{sec:similarity}

The graph appeared to suggest five main classes of coordination geometries: (1) \textit{spheroidals}, characterised by their high sphericity; (2) \textit{ellipsoidals}, characterised by their high moment of intertia; (3) \textit{cuboidals}, characterised by their cubic core; (4) \textit{bipyramidals}, characterised by their bipyramidal form; and (5) the \textit{tetrahedral}, a class on its own. Statistics for these classes are given in Table \ref{tab:averages}. Augmenting the classes with the clusters from the binary tree resulted in the taxonomy depicted in Fig. \ref{fig:graph}.

Upon close inspection, we found the positions of individual geometries within classes to exhibit consistency with the overall taxonomy. For example, the proximity of TET to the cuboidals makes sense given that its $4$ bonds correspond to $4$ of the $8$ edges of a cube. Likewise, the proximity of PBP and HBP to the spheroidals is justified by the fact that they are the biypramidal geometries of highest sphericity, as is the proximity of BTP and SDS to the spheroidals, given that they are the trigonal prismatics of highest sphericity.

\begin{table}[b]
    \centering
    \begin{ruledtabular}
    \begin{tabular}{lccc}
    Class          & $\Psi$ & $I/k$  & $\tau_E$ \\ \midrule
    1. spheroidal  & $0.90$ & $1.09$ & $-0.74$  \\
    2. ellipsoidal & $0.84$ & $1.28$ & $-0.95$  \\ 
    3. bipyramidal & $0.83$ & $1.04$ & $-0.75$  \\
    4. cuboidal    & $0.85$ & $1.12$ & $-0.73$  \\
    5. tetrahedral & $0.67$ & $1.00$ & $-1.36$  \\
    \end{tabular}
    \end{ruledtabular}
    \caption{Class-average parameters: sphericity $\Psi$, moment of inertia per neighbour $I/k$, and extracopular typicality $\tau_E$ (see Appendix \ref{appx:parameters} for details on the former two quantities).}
    \label{tab:averages}
\end{table}

\subsubsection{Typicality}
\label{sec:typicality}

An important finding of the original work on $E$ was that ICO is the maximally ordered coordination geometry among those commonly encountered and that FCC is the most ordered such lattice \cite{camkiran_2022}. An interesting follow-up question that can be asked is how `typical' a given coordination geometry is relative to others. To answer this question, we define the \textit{extracopular typicality} $\tau_E(g)$ of a geometry $g$ as its negative extracopularity distance from the centroid $\avg{h}$ of all three-dimensional coordination geometries,
\begin{equation}
    \tau_E(g) -d_E(g,\avg{h}).
\end{equation}

We computed typicalities numerically using the $8$-dimensional coordinates obtained through MDS, taking $\avg{h}$ to be the componentwise average over all $22$ geometries under study. We found TET to be the geometry of lowest typicality, which is not surprising given its anomalously low coordination number and bond angle count. Meanwhile, HBP was found to be the geometry of highest typicality, which also makes sense given that it is quite usual in both regards. \mbox{Table \ref{tab:averages}} provides class averages on typicalities; \mbox{Table \ref{tab:cegs}} gives values for individual geometries. 

\begin{figure}[t]
    \centering
    \begin{tabular}{cc}
         \rotatebox[origin=c]{90}{\hspace{5mm} typicality $[\tau_E(g)]$} &  \begin{tabular}{c} \includegraphics[width=0.88\linewidth]{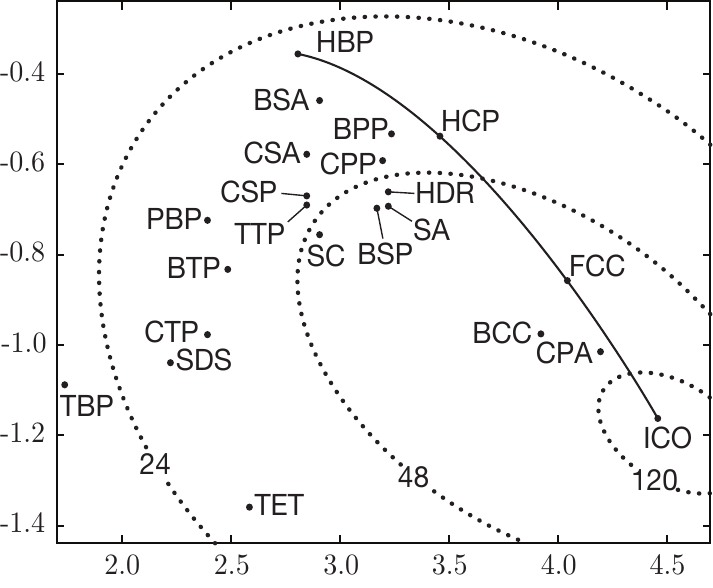} \end{tabular} \\
          & \hspace{8mm} order $(E_g)$
    \end{tabular}
    \caption{Typicality versus order. The number on each dotted ellipse indicates the maximum point group order of any geometry in the region between that ellipse and the one smaller. The solid curve indicates the apparent frontier.}
    \label{fig:scatter}
\end{figure}

To investigate the interplay between typicality and order, we produced a scatter plot of $\tau_E(g)$ versus $E_{g}$, presented in Fig. \ref{fig:scatter}. The plot appeared to evidence a statistical regularity with respect to point symmetry. In particular, it was possible to construct nested ellipses of decreasing maximum point group order. We also observed an apparent `frontier' of geometries with high typicality for a given level of order.

At the highest-order end of the frontier we found ICO, the coordination geometry that is locally preferred in many monodisperse systems, notably simple liquids \cite{frank_1952}, metallic glasses \cite{ding_2014}, and hard-sphere packings \cite{clarke_1993}. Further along the frontier were FCC and HCP, which describe the structural tendency of a wide variety systems \cite{heitkam_2012}. And at the lowest-order end of the frontier was HBP, the coordination geometry of particles arranged in a simple hexagonal lattice---the ground state structure for certain `core-softened' potentials \cite{quigley_2005,rechtsman_2006}.

It is also worth remarking on the geometries found near (but not on) the frontier. These include BSA, which is known to be locally preferred by the Kob-Andersen mixture \cite{kob_1995,malins_2013_a}; BPP, which closely resembles HCP; BCC, which corresponds to the lattice of the same name; HDR, a subset of BCC; and CPA, a subset of ICO. 

Lastly, we note the geometries observed furthest from the frontier, TBP and TET, both of which are associated with materials of highly anomalous properties, such as graphite, diamond, water, and silica. \cite{shi_2018}.

\section{Discussion}
\label{sec:discussion}

In essence, this work advances a redundancy-based perspective on coordination geometries as an alternative to traditional symmetry-based thinking. Point symmetry is itself of course a kind of very useful redundancy \cite{bradley_2010}. However, there are at least two ways in which it is unsuited to the study of coordination geometries. Firstly, it is degenerate to the extent of not being able to distinguish between basic crystal structures (e.g. FCC, BCC, and SC). Secondly, small changes to a given geometry, such as the removal of a neighbour, can lead to large differences in its symmetry (cf. ICO and CPA). In assigning positive distances to all $231$ pairs of geometries herein studied (Fig. \ref{fig:heatmap-full}), our perspective does not appear to evidence any degeneracies. And in suggesting a geometrically meaningful taxonomy (Fig. \ref{fig:graph}), it does not appear to avoid degeneracies at the expense of informativeness.

The central result of this work is our metric model of the space of three-dimensional coordination geometries. It is conceivable that the metric captures, if only partially, the energetic or entropic cost that would be associated with a transformation from one geometry to another \cite{plessow_2020}. Certainly, the well-known pathways of Bain \cite{bain_1924} and Burgers \cite{burgers_1934} are consistent with this idea, as the geometries they involve lie in close proximity to one another in our model. Further work would be needed to explore this hypothesis.

In our previous work, we observed that a pairwise informational redundancy in the bonds formed by a particle underlies orientational order \cite{camkiran_2022}. In the present work, we find that the same phenomenon underlies orientational dissimilarity and orientational typicality. It would thus appear that this single phenomenon, which we call extracopularity, is able to reconcile three otherwise distinct, orientational aspects of local structure---order, dissimilarity, and typicality.

In addition to the theoretical implications discussed above, our results also have implications on the practical problem of local structural indication \cite{stukowski_2012,tanaka_2019}. The present work asserts a set of axioms on the topology of the space of coordination geometries to arrive at an optimal method of discretizing bond angles. Bond angle discretization makes it possible to compute one-particle $E$ with greater precision than by the workaround discussed in Ref. \cite{camkiran_2022}. This greatly increases the range of structures that are possible to resolve with $E$ \textit{in silico}.


\begin{acknowledgments}
The authors would like to thank Al{\'a}n Aspuru-Guzik and Chandra Veer Singh for their helpful discussions.
\end{acknowledgments}

\appendix
\section{Commonly encountered geometries}
\label{appx:cegs}
\begin{figure*}
    \centering
    \includegraphics[width=\linewidth]{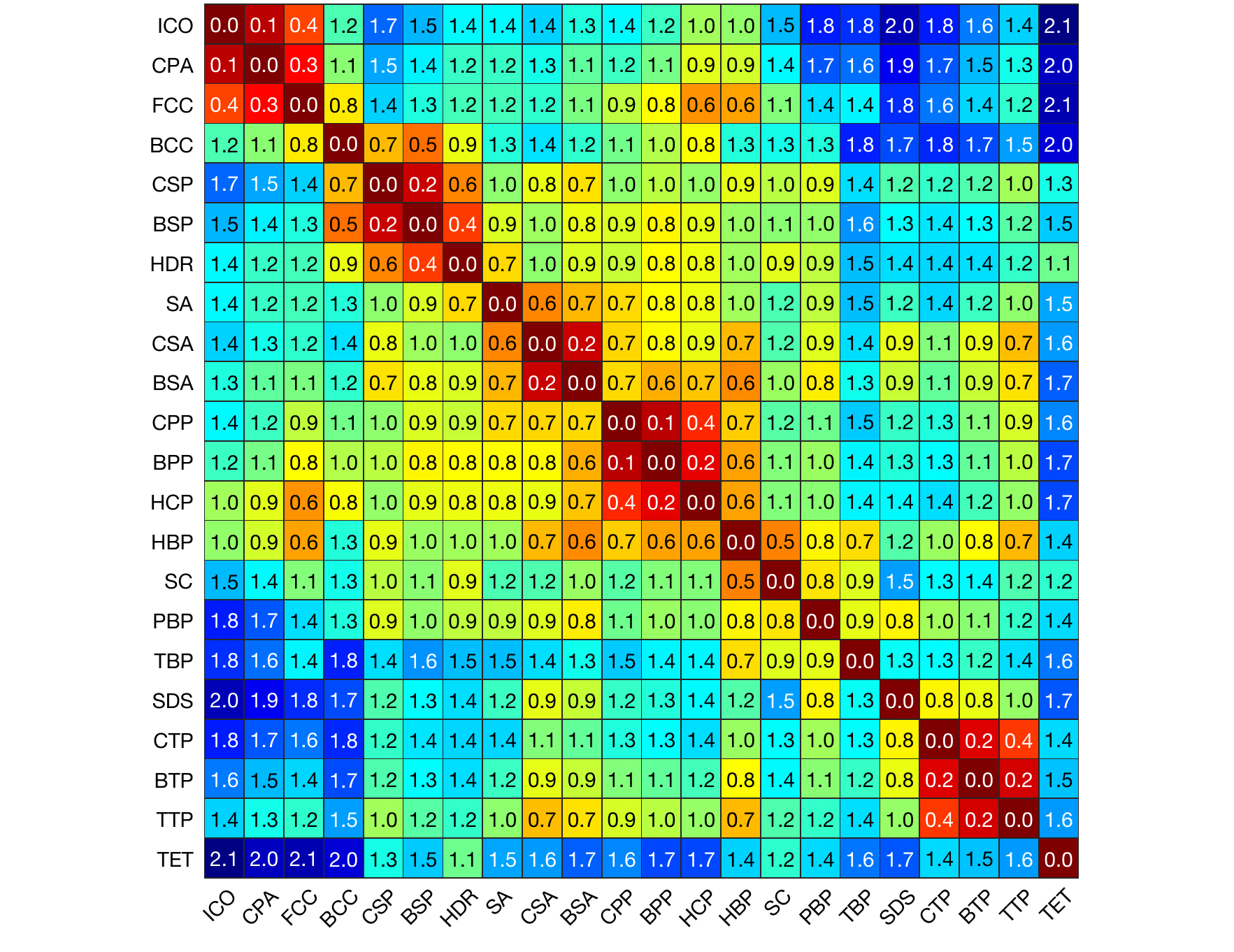}
    \caption{Heatmap of extracopularity distances for all $22$ commonly encountered geometries. All values are rounded to one decimal place. Rows/columns are ordered as per \mbox{Ref. \cite{bar-joseph_2001}}.}
    \label{fig:heatmap-full}
\end{figure*}

Table \ref{tab:cegs} lists $22$ of the most commonly encountered geometries of coordination. These include $11$ zero-lone-pair molecular geometries predicted by valence shell electron pair repulsion theory \cite{gillespie_2008} and $8$ solutions of the Thomson problem \cite{thomson_1904}, the latter of which also happen to be minimum-energy sphere packings for $k \leq 12$ \cite{atiyah_2003}. Mathematically, these correspond to the first $4$ Platonic solids, all $8$ strictly convex deltahedra, $12$ capped (anti)prisms, $4$ regular bipyramids, $2$ circumscribable bicupolae, and the rhombic dodecahedron (a Catalan solid). Fig \ref{fig:heatmap-full} depicts their extracopularity distances.

\begin{table*}
    \centering
    \begin{ruledtabular}
    \begin{tabular}{lllrrrrrr} \multicolumn{3}{c}{\centered{Description}} &  \multicolumn{6}{c}{Parameters} \\ \cmidrule{1-3} 
    \cmidrule{4-9} \multicolumn{1}{l}{Abbrev.} & \multicolumn{1}{c}{Coordination geometry} & \multicolumn{1}{c}{Polyhedral classification} & \multicolumn{1}{c}{$k$} & \multicolumn{1}{c}{$m$} & \multicolumn{1}{c}{$E$} & \multicolumn{1}{c}{$\tau_E(g)$} & \multicolumn{1}{c}{$\Psi$} & \multicolumn{1}{c}{$I/k$} \\\midrule
    TBP & Trigonal bipyramidal & Deltahedral, bipyramidal & $5$ & $3$ & $1.737$ & $-1.0880$ & 0.7563	& 1.14 \\
    SDS & Snub disphenoidal\footnote{Often called trigonal dodecahedral} & Deltahedral & $8$ & $6$ & $2.222$ & $-1.0392$ & 0.8543 & 1.31 \\
    PBP & Pentagonal bipyramidal & Deltahedral, bipyramidal &  $7$ & $4$ & $2.392$ & $-0.7239$ & 0.8696 &	1.00 \\
    CTP & Capped trigonal prismatic & Prismatic & $7$ & $4$ & $2.392$ & $-0.9770$ & 0.8025 & 1.33 \\
    BTP & Bicapped trigonal prismatic & Prismatic & $8$ & $5$ & $2.485$ & $-0.8325$ & 0.8630 & 1.19 \\
    TET & Regular tetrahedral & Platonic, deltahedral & $4$  & $1$ & $2.585$ & $-1.3589$ & 0.6711	&	1.00 \\
    HBP & Hexagonal bipyramidal & Bipyramidal &  $8$ & $4$ & $2.807$ & $-0.3556$ & 0.8630 &	1.00\\ 
    CSA & Capped square antiprismatic & Antiprismatic & $9$ & $5$ & $2.848$ & $-0.5778$ & 0.8778	&	1.24 \\
    CSP & Capped square prismatic & Prismatic & $9$ & $5$ & $2.848$ & $-0.6698$ & 0.8272	& 1.28 \\
    TTP & Tricapped trigonal prismatic & Prismatic, deltahedral & $9$ & $5$ & $2.848$ & $-0.6901$ &	0.9062	&	1.00 \\
    SC & Regular octahedral\footnote{Square bipyramidal} & Platonic, deltahedral, bipyramidal & $6$ & $2$ & $2.907$ & $-0.7558$ & 0.8456	&	1.00  \\
    BSA & Bicapped square antiprismatic & Deltahedral, antiprismatic & $10$ & $6$ & $2.907$ & $-0.4588$ & 0.8853	&	1.18 \\
    BSP & Bicapped square prismatic & Prismatic & $10$ & $5$ & $3.170$ & $-0.6972$ & 0.8579	&	1.07 \\
    CPP & Capped pentagonal prismatic & Prismatic & $11$ & $6$ & $3.196$ & $-0.5918$ & 0.8695	&	1.20 \\
    SA & Square antiprismatic & Antiprismatic &  $8$ & $3$ & $3.222$ & $-0.6928$ & 0.8595	&	1.00   \\
    HDR & Regular hexahedral\footnote{Square prismatic, cubic} & Platonic, prismatic & $8$ & $3$ & $3.222$ & $-0.6609$ & 0.8060	&	1.00 \\
    BPP & Bicapped pentagonal prismatic & Prismatic & $12$ & $7$ & $3.237$ & $-0.5326$ & 0.9095	&	1.00 \\
    HCP & Anticuboctahedral\footnote{Triangular orthobicupolar} & Bicupolar & $12$ & $6$ & $3.459$ & $-0.5377$ & 0.9050	&	1.00 \\
    BCC & Rhombic dodecahedral & Catalan & $14$ & $6$ & $3.923$ & $-0.9754$ & 0.9047	&	1.14  \\
    FCC & Cuboctahedral\footnote{Triangular gyrobicupolar} & Bicupolar & $12$ & $4$ & $4.044$ & $-0.8574$ &	0.9050	&	1.00 \\
    CPA & Capped pentagonal antiprismatic & Antiprismatic & $11$ & $3$ & $4.196$ & $-1.0148$ & 0.8967	& 1.20  \\
    ICO & Regular icosahedral\footnote{Bicapped pentagonal antiprismatic} & Platonic, deltahedral, antiprismatic & $12$ & $3$ & $4.459$ & $-1.1625$ &	0.9393	&	1.00 \\
    \end{tabular}
    \end{ruledtabular}
    \caption{List of commonly encountered geometries in order of increasing $E$, rounded to three decimal places. See \mbox{Appendix \ref{appx:parameters}} for details on sphericity $\Psi$ and moment of inertia per neighbour $I/k$.}
    \label{tab:cegs}
\end{table*}

\section{A technicality of fixed discretization}
\label{appx:discretization}

One of the inevitabilities of fixed bond angle discretization is the confusion of angles that are close yet unequal even in the ideal form of a coordination geometry. Among the geometries studied here, this issue is observed to afflict those of type (X)PP, (X)SA, (X)TP, and SDS. For the analysis described in \mbox{Sec. \ref{sec:topology}}, we corrected their bond angle counts as follows. Let $f_{g}(\theta)$ denote the number of close yet unequal angles that are discretized to the (same) angle $\theta$ for a coordination geometry $g$. Then, we define the corrected cardinality of $\Theta_{gh}$ as
\begin{equation}
    \abs{\Theta_{gh}}' = \sum_{\theta \in \Theta_{gh}} \max \{ f_{g}(\theta), f_{h}(\theta) \}.
\end{equation}

\section{Miscellaneous parameters}
\label{appx:parameters}

\subsection{Sphericity}

The \textit{sphericity} $\Psi$ of a particle with a given three-dimensional coordination geometry is defined as the fraction of the surface area of a sphere with the same volume $V$ as the geometry to the surface area $A$ of the geometry itself,
\begin{equation}
    \Psi = \frac{\pi^{1/3}(6V)^{2/3}}{A}
\end{equation}
\cite{wadell_1935}. By the isoperimetric inequality, we have $0 < \Psi < 1$.

\subsection{Moment of inertia per neighbour}

The \textit{moment of inertia per neighbour} $I/k$ of a particle with a given coordination geometry can be computed as follows:
\begin{enumerate}
    \item Calculate the centroid $c$ of the particle's neighbourhood as the average position of its neighbours, $c = \avg q$.
    \item Determine the Euclidean distance $\ell_q$ of each neighbour $q$ from the centroid, $\ell_q = \abs{c-q}$.
    \item Take the sum of the square of these distances to get the moment of inertia, $I = \sum_q \ell_q^2 $.
    \item Divide $I$ by $k$ to get the per-neighbour value.
\end{enumerate}

\bibliography{main.bib}


\end{document}